\documentstyle[11pt,aaspp4,epsf]{article}
\begin{document}

\title{Kinematics and Structure of the Starburst 
Galaxy NGC~7673} 

\def\km{{\rm\,km}}
\def\kms{{\rm\,km\,s^{-1}}}
\def\kpc{{\rm\,kpc}}
\def\mpc{{\rm\,Mpc}}
\def\msun{{\rm\,M_\odot}}
\def\lsun{{\rm\,L_\odot}}
\def\rsun{{\rm\,R_\odot}}
\def\pc{{\rm\,pc}}
\def\cm{{\rm\,cm}}
\def\yr{{\rm\,yr}}
\def\au{{\rm\,AU}}
\def\g{{\rm\,g}}
\def\om{\Omega_0}
\def\ltsima{$\; \buildrel < \over \sim \;$}
\def\simlt{\lower.5ex\hbox{\ltsima}}
\def\gtsima{$\; \buildrel > \over \sim \;$}
\def\simgt{\lower.5ex\hbox{\gtsima}}
\tighten

\author{N. L. Homeier and J. S. Gallagher}
\affil{Department of Astronomy, University of Wisconsin-Madison}
\authoraddr{475 N. Charter St., Madison, WI 53706, USA}

\begin{abstract}

The morphology and kinematics of the luminous blue starburst galaxy
NGC~7673 are explored using the WIYN 3.5m telescope.  Signs of a past
kinematic disturbance are detected in the outer galaxy; the most
notable feature is a luminous ripple located 1.55 arc minutes from the
center of NGC~7673.  Sub-arc second imaging in B and R filters also
reveals red dust lanes and blue star clusters that delineate spiral
arms in the bright inner disk and narrow band H$\alpha$ imaging shows that
the luminous star clusters are associated with giant H~II
regions. The H$\alpha$ kinematics measured with echelle imaging spectroscopy
using the WIYN DensePak fiber array imply that these H~II regions are
confined to a smoothly rotating disk.  The velocity dispersion in
ionized gas in the disk is $\sigma \sim$ 24~km~s$^{-1}$, which sets an upper 
bound on the dispersion of young stellar populations.. Broad
emission components with $\sigma \sim$ 63~km~s$^{-1}$ found in some
regions are likely produced by mechanical power supplied by
massive, young stars; a violent starburst is occurring in a kinematically calm
disk. Although the asymmetric outer features point to a merger or
interaction as the starburst trigger, the inner disk structure constrains
the strength of the event to the scale of a minor merger or weak
interaction that occurred at least an outer disk dynamical time scale
in the past.
\end{abstract}


\section{Introduction}

High surface brightness blue galaxies have been a curiosity in samples
of nearby galaxies for more than 40 years (e.g. \cite{Z57},
\cite{S63}).  While these types of compact galaxies were often
excluded from traditional samples of galaxies that chose objects on
the basis of angular size (see \cite{A66}), they can be readily found
in surveys that select for blue colors or strong optical emission
lines. Follow-up studies to the Markarian survey, for example, have
shown that these high surface brightness galaxies consist of systems
with AGNs and/or with extraordinarily high star formation rates (SFRs), 
i.e. the starburst galaxies (e.g., \cite{S1970}, \cite{H77}). 
Recently interest in blue high SFR galaxies has been stimulated by their
spectroscopic and morphological similarities to classes of galaxies
which are commonly found at moderate to high redshifts (see
\cite{GHB89}, \cite{G1990}, \cite{CHS1995}, \cite{E1997}, \cite{Getal98}). 
Hence, studying 
nearby starbursts can further our understanding of rapidly evolving galaxies
observed at moderate to large lookback times. 

High surface brightness starburst galaxies with blue optical colors 
comprise about 5\% of the local
galaxy field population with optical luminosities of $L_B$ \gtsima 
3$\times$10$^9 \lsun$ (e.g. \cite{H77}). The origins, structures, and
evolution of these luminous blue galaxies (LBGs), however, are not
well-understood. We therefore have undertaken a new study of nearby
LBGs with the objectives of obtaining better information on their
structures, characterizing spatial patterns of star formation,
and measuring internal kinematics. The dynamics of these systems are
particularly important for predicting their subsequent evolution. 
Many of these galaxies are optically peculiar at the moment we see them, 
due to enhanced SFRs, but their appearance at later times may be determined 
by the present locations and kinematics of their stars and gas, which in 
turn influence future star formation processes.

In this paper we report results from a study of emission line kinematics
and optical imaging  with the WIYN\footnote{The WIYN Observatory
is a joint facility of the University of Wisconsin-Madison, Indiana
University, Yale University, and the National Optical Astronomy
Observatories.} 3.5m telescope in the archetypal ``clumpy irregular'' galaxy 
NGC~7673\footnote{The RC2 (\cite{DDC1976}) and Huchra (1977) classified
NGC 7673 as a disturbed spiral, while Casini \& Heidmann noted that large
star forming clumps in an envelope are characteristic of luminous UV bright
galaxies which they called 'clumpy irregulars', of which NGC 7673 is a 
prime example.} (\cite{CH1976}, \cite{TH1986}, \cite{BCH1982}). This galaxy 
has 
a highly disturbed optical structure, which leads naturally to the question 
of its dynamical state. Is this a disrupted system, or an example of unusual
star formation processes within a dynamically normal disk galaxy? 
A second paper in this series focuses on star formation patterns and timescales
from  high angular resolution imaging of NGC~7673 obtained with the Wide 
Field Planetary Camera 2 on the {\it Hubble Space Telescope} (\cite{Getal99}).

\section{Properties of NGC~7673}

The unusual optical structure of NGC~7673 ($=$ IV~Zw~149, Mrk~325) was
initially described in terms of the presence of multiple bright knots (e.g.,
\cite{ML1971}, \cite{BK1975}). High quality ground-based images,
primarily obtained by J. Heidmann and his collaborators, indicated
this galaxy consists of giant star-forming clumps embedded in a
diffuse halo, and thus NGC~7673 was classified as a clumpy irregular
galaxy (\cite{CH1976}, \cite{Cetal1982}). The superb ground-based
images of Coupinot et al. (1982) also showed the spiral
pattern that was noted in the RC2 and by Huchra (1977), 
which becomes clearer in images taken with WFPC2 (\cite{Getal99}). The 
hyperactive star formation within the clumps deduced by
Heidmann and collaborators was confirmed by Gallego et al. (1996); their 
data indicate that L(H$\alpha$) for the three largest is 2-8$\times$10$^8 
\lsun$. These are each comparable to the total L(H$\alpha$) seen in many 
starburst galaxy nuclei.

Emission line kinematics were measured in NGC~7673 by Duflot-Augarde
\& Alloin (1982; hereafter DA), who found a remarkably constant
velocity across the galaxy. The lack of
internal velocity spread led DA to conclude that NGC~7673 is rather
quiescent, and they therefore rejected a merger model, but noted that
an interaction with its neighboring galaxy, NGC~7677 (Mrk 326) at a
similar radial velocity (3554 $\kms$ versus 3405 $\kms$), is
possible, as suggested by Casini \& Heidmann (1976). Further 
observations of the emission lines in NGC~7673 by
Taniguchi \& Tamura (1987) taken at higher spectral resolution showed
the presence of a broad line component beneath the narrow emission
line in clump B, the northwestern clump. They suggested that this broad 
feature, for which
they measured a full width zero intensity (FWZI) of 420$\kms$, is due 
to mass motions powered by supernova activity.

NGC 7673 appears to be a member of a small galaxy group that is relatively 
isolated, possibly located on the outskirts
of the Pegasus I cluster and in front of the Pisces-Perseus chain.
The immediate environment of the NGC~7673-7 pair was mapped in the HI
21~cm line by Nordgren {\it et al.} (1997), who find an H~I mass of 
3.6 $\pm$ 0.5 x 10$^{9}$ M$_{\odot}$ for NGC~7673. Both galaxies contain
extended HI disks; NGC~7673 is apparently nearly face-on with a modest
HI-asymmetry to the west. NGC~7677 contains a small outer HI
irregularity that points towards its neighbor, and is at nearly its
radial velocity. Thus while an interaction is not excluded, there is
no striking evidence for an on-going event in this system.

In this paper we assume H$_0 =$75$\kms$ Mpc$^{-1}$. The recession
velocity of 3405~$\kms$ for NGC~7673 then implies a distance of
46~Mpc, and a linear distance of 222 pc per arcsec. Other parameters
in the NASA/IPAC Extraglactic Database\footnote{The NASA/IPAC
Extragalactic Database (NED) is operated by the Jet Propulsion
Laboratory, California Institute of Technology, under contract with
the National Aeronautics and Space Administration.} are B$^0_T =$12.81
implying M$_B=-$20 and blue global colors of (B$-$V)$_0=$0.34 and
(U$-$B)$_0=-$0.38.

Our observations and reductions are detailed in \S 2. In \S 3 we
present our results, which are discussed in \S 4, while \S 5 contains
a summary and conclusions.

\section{Observations \& Reductions}

All observations were made with the WIYN 3.5m, f/6.5 telescope at the 
Kitt Peak National
Observatory. All data reduction was performed with IRAF\footnote{IRAF
is provided by the courtesy of the National Optical Observatories,
which are operated by the Association of Universities for Research in
Astronomy, Inc., under cooperative agreement with the US National
Science Foundation.}.

\subsection{Optical Imaging}

Images were obtained using a Tektronics 2048 x 2048 pixel CCD with a field of 
6.7 x 6.7 arcmin and a scale of 0.195 arcsec per pixel. On November
13, 1996 a 500 s exposure was taken with a Harris R filter, and a 700 s
exposure with a Harris B filter. The images were corrected for bias
and zero level and flat-fielded with the `ccdproc' task; cosmic rays
were not removed.  The seeing in these images is 0.''8.

Narrow band imaging was undertaken on July 24, 1997, a non-photometric
night with moderately poor seeing. We were able to get 2 exposures
taken with a redshifted H$\alpha$ filter, 1000 s and 631 s
(cloud-dodging), and a 300 s exposure taken with a Harris R filter. The
redshifted H$\alpha$ filter is centered at 6618 \AA$\;$ and has a
width of 72 \AA. Thus our H$\alpha$ narrow band images also include
emission from [N~II].

These H$\alpha$ and R images were processed in the same way as the
earlier data, except that we removed cosmic rays from the pair of
narrow band images.  Both the combined H$\alpha$ and R image were
scaled in such a way that subtracting the R image from the combined
H$\alpha$ image canceled out field stars, producing a
continuum-subtracted H$\alpha$ image. The seeing in this final
H$\alpha$ image is about 1''.4.

\subsection{DensePak Optical Spectroscopy}

On October 23, 1997, two 1500s exposures were taken with the fiber array 
DensePak (see \cite{BSH98}) on the WIYN Nasmyth port. DensePak is a
fixed array of 91 red-optimized fibers arranged in a 7 x 13
rectangle. Each fiber has a 3'' diameter, and the fiber centers are
each separated by 4'' to form a 30'' x 45'' array. The fiber cable
feeds the Bench Spectrograph, which for these observations was
configured with a 316 line mm$^{-1}$ echelle grating and a grating
angle of 62.581 degrees.  We used the Bench Spectrograph camera and
the 2048 x 2048 T2kC CCD detector with 24$\mu$m per pixel, for a
wavelength coverage of approximately 6300 to 6730 \AA\ with a reciprocal
dispersion of 0.13 \AA\ per CCD pixel.

A DensePak schematic is shown in Figure 1. There are 4 fibers placed
off the corners of the array: numbers 8, 32, 68, and 90. We were able
to use these for the intended purpose of sky subtraction because these
fibers were not covering the H$\alpha$ emitting regions of the galaxy,
so we did not have to move the telescope for a separate sky
exposure. Also note the gaps in numbering; 23 and 28 have been
skipped. Fibers 46 and 59 are dead.

Bias frames and dome flats were taken at the beginning and end of the
night and combined during reduction with the IRAF task `dohydra'. For
wavelength calibration, comparison spectra were taken with a ThAr lamp
after each exposure. The 4 spectra from the sky fibers were combined
to make a single sky spectrum and subtracted from each of the other
fibers as part of the 'dohydra' procedure. Combining the 2 final
zeroed, flat-fielded, sky-subtracted spectra was accomplished with
`scombine' with reject=`crreject' to remove cosmic rays.

To estimate the error in our measurements, Gaussian profiles were fit
to the night sky lines in the combined sky spectrum with the IRAF task
`splot'. The observed wavelength of the H$\alpha$ line is close to
6638 \AA$\;$ over the entire galaxy, so comparison night sky lines
were chosen near this wavelength. The FWHM resolution determined in
this way is 32 $\kms$. The statistical uncertainty in central
wavelength for a bright sky line refers to the scatter about the
absolute wavelength, and is $\leq$ 0.05 \AA, which corresponds to 2.3
$\kms$ at 6638 \AA. However, in some cases the complex profiles or low
signal-to-noise ratios of the H$\alpha$ emission lines makes the
determination of accurate radial velocities difficult, yielding a greater
uncertainty.
    
Gaussian emission profiles were assumed and fitting done with the IRAF
task 'splot'.

\section{Results}

\subsection{Imaging}

The deep R-band image is shown in Figure 2a.; an unsharp mask of the
same image is shown in Figure 2b. Surrounding the bright central disk
of NGC~7673 are several wispy, low-surface brightness features. In Figure
2b one can see a broad 'bridge' extending to the northwest connecting
the inner disk to the first arc at 36'' (8 kpc) from the galaxy
center. The next arc, or ripple, can be seen in Figure 2a, located to
the east at 52'' (11.5 kpc) from the center. The most striking ripple
is located to the west at 1'.55 (21 kpc) from the center of the bright
optical disk. This sharply defined arc was first reported by Dettmar
et al. (1984), who suggested this feature either could be stars or an
emission line region. In Figure 2b. there are several extensions to
the main disk, most noticeablely two on the eastern edge and one in
the southwest corner. The linear feature north of NGC 7673 is a background
galaxy, which is confirmed by its extreme redness in the B$-$R image and the
Gallagher et al. (1999) WFPC2 images.

The R-band image is shown again in Figure 2c., with a stretch that
shows the structure of the inner disk, including the central bar. That the 
light gray smudge in the SE corner is a background galaxy is confirmed from 
its B$-$R color and the WFPC2 images which reveal it to be a distant 
(and beautiful) spiral galaxy. A B$-$R color map
is shown in Figure 2d. The B and R images have not been
photometrically calibrated, so the color map is a B/R ratio map after
standard reduction and sky subtraction. The spiral nature of the inner
disk is clearly visible; there are prominent red dust lanes and bright
blue clusters in this image. Unfortunately, we were not able to get
color information on the outer ripples; they are too faint for
reliable color measurements from these data. However, the ripples are
present in both B and R images with comparable intensities, indicating
that they are composed of stars.

A continuum subtracted H$\alpha$ image is shown in
Figure 3. There are 3 main H~II regions: the central 'nucleus', the large clump
to the NE, and a smaller clump situated between the two. These
correspond to clumps A, B, and C, respectively, as referred to by DA
and Taniguchi \& Tamura (1987). In the SE corner there is a fainter
H$\alpha$ ring, where it is possible to identify smaller H~II
regions. A bright cluster of stars seen in the B and R -band images is
located on the northern side of this crooked 'arm', nestled between
edges of ionized gas visible in the H$\alpha$ image. The gas directly
between us and the cluster does not appear in the H$\alpha$ image, but
the gas in the surrounding galaxy plane is brightly lit. The diffuse
outer features are not detected in our narrow band H$\alpha$ image,
consistent with our interpretation that they are starlight.

\subsection{Spectroscopy}

We performed a kinematic study of NGC~7673 using the DensePak fiber array
and the H$\alpha$ emission line redshifted to near 6638
\AA. To find fiber placement on the galaxy, the narrow band
H$\alpha$ image, Figure 3, is compared to the DensePak map with flux in 
gray-scale, shown in Figure 4. The approximate placement of the DensePak array
is marked with a box in Figure 3. Figure 4 was produced with the
IRAF task 'sbands', taking the flux in a 10~\AA$\;$ wide band centered
on 6638~\AA$\;$ and using 6~\AA$\;$ wide bands to the blue and red for
continuum subtraction. The fiber diameter of 3'' covers a linear
distance of approximately 650 pc at NGC~7673, but we were not able to
cover all of the main body of the galaxy with our single DensePak
pointing.

\subsubsection{Emission Profiles}

No emission was detected in 29 of the 89 working fibers; the fiber
illumination pattern is consistent with our H$\alpha$ image. Of the
remaining 60 fibers, 32 had emission fit by a single Gaussian, and 28
were deconvolved into two Gaussian components, one broad and one
narrow. It became clear that deconvolution was necessary when a single
component fit left large, broad, asymmetric shoulders of emission;
most of these broad lines are found in the northern half of the
galaxy. Sample spectra are displayed in Figure 5.

Fibers on the southern part of the galaxy were fit by a single
Gaussian, but in most cases small shoulders, or wings, of emission 
often remained. These wings are similar in shape to those in the middle of 
the array, but much less pronounced. Even in
spectra with similar peak emission, the wings were smaller in the
southern part of the galaxy. A deconvolution of these spectra into two 
components improved the fit, but with one or both of the components having a
 FWHM at or below the
instrumental resolution. Therefore, these spectra were fit with a
single Gaussian. Figure 6 is a plot of FWHM, corrected for the instrumental
profile (FWHM$^{2}$=(FWHM$_{obs}$)$^{2}$-(32 km s$^{-1}$)$^{2}$), versus 
position, which
shows the distinct widths of the two components, and some scatter, but 
no systematic variation $>$ 10 km s$^{-1}$ in the FWHM of the narrow 
emission component over the galaxy.

Some spectra show what appear to be multiple peaks, perhaps due to
emission from two or more individual H~II regions, or splitting from 
expanding shells or bubbles. Since most of these spectra have low S/N, a
single-Gaussian fit was forced to recover a mean central velocity and
a FWHM of dubious meaning. Such profiles were observed in 12 fibers; 6 had
FWHMs consistent with the narrow component, and 6 had FWHM $>$ 
90 km s$^{-1}$.

\subsubsection{Velocity Field}

The velocity field as measured by the heliocentric velocities of the
narrow components is shown in Figure 7, where the velocity is in km s$^{-1}$
and grayscale shading is used to help visualize the velocity field. Filled 
polygons indicate
spectra with no narrow emission component. Fibers with broad emission but not
narrow are thus represented by filled polygons; this includes fibers 55, 62,
72, 79, and 81. Fibers which were fit by both a narrow and a broad Gaussian 
are marked with a small cross above the velocity. 

We find the narrow component velocity field to be fairly
constant, with a radial velocity difference of only $\sim$ 60 $\kms$ 
across the galaxy
regions covered by DensePak (the western side of the galaxy was not
sampled). The maximum velocity is 3444 km s$^{-1}$ in the SW corner,
and falls to 3403 $\kms$ at the northern edge. The minimum, 3392
$\kms$, occurs near the nuclear region, and the recession velocity of
the nucleus itself is $\sim$ 3407 $\kms$. These results agree with
previous measurements by DA. This also agrees with the H I data
(\cite{Netal1997} which shows the peak emission to be at a
heliocentric velocity of 3410 $\kms$. The precision of the DensePak
absolute radial velocities is similar to that obtained from HI 21-cm
line measurements.

The shallow velocity gradient across NGC~7673 indicates that the H~II
regions are confined to a dynamically cold rotating disk that
is nearly face-on. The optical and H~I morphology of NGC~7673 also suggest
the presence of a disk, whose existence is confirmed by these emission
line measurements showing rotation about its barred nuclear region.
Although the disk shows signs of perturbation, such as copious star
formation and a disturbed spiral pattern, it has survived the event
which triggered the starburst and appears to be rotating smoothly.

The velocity gradient measured by optical emission lines (here, and by
DA), although modest due to a small angle of inclination, differs
from that of the H I disk (\cite{Netal1997}). The western edge of the
optical disk is approaching and the eastern receding, but the H~I
velocity gradient is in the opposite sense. This may be a signature
that moderate-scale kinematic disturbances remain in the disk of
NGC~7673, and high angular resolution H~I maps would provide a useful
diagnostic of the state of this galaxy.

\section{Discussion}

\subsection{Kinematics}

In regions of low flux, the emission line profiles are usually Gaussian 
in shape; however, as previously mentioned, 
12 fibers contain spectra with large asymmetries. 
Over the central regions of the galaxy where the flux is highest, we find 
H$\alpha$ emission
lines with a relatively narrow core and broad wings, very similar to the 
integrated profiles of giant H~II regions found in many previous studies 
(\cite{SW1970}, \cite{GH1983}, \cite{SB1984}, \cite{AR1986}, 
\cite{RAJ1986}, \cite{AR1988}, 
\cite{Retal1998}). Some of these emission lines are well fit by a Voigt
profile, but most have asymmetric shoulders, and thus are best fit with
two Gaussian components. We therefore assume that the emission line profiles
which resemble Voigt profile are a combination of two Gaussian components with 
negligible velocity offset, and fit all such profiles with two Gaussians, 
leaving the central wavelengths and line widths free to vary. Taking the 
emission component(s) in each fiber, we find the mean FWHM of each
kinematic component is 56 and 149 $\kms$,
respectively ($\sigma$ $\sim$ 24 km s$^{-1}$, $\sigma$ $\sim$ 63 km s$^{-1}$). 
Fibers with emission lines which were fit with two profiles
are marked with a small cross in Figure 7. Most of the broad
components are at a slightly lower velocity relative to the narrow
line in the same spectrum, but never by more than 10 $\kms$. Taniguchi
\& Tamura (1987) also found a narrow and a broad component in the
H$\alpha$ emission line from clump B; they 
measured the broad
component as having a FWHM $>$ 200$\kms$, and found it was redshifted 
by 13 $\kms$ relative to the narrow component of FWHM $\sim$ 70 $\kms$.

The mean FWHM of the narrow component is much broader than emission 
from typical H~II regions in our own galaxy (typical FWHM $\sim$ 25$\kms$, see
\cite{M58}, \cite{FTD90}). 
The origin of ionized gas with supersonic velocity dispersions is still a
subject of debate, and thus a generally accepted model for this
phenomenon does not yet exist. An L-$\sigma$ (luminosity-velocity dispersion) 
and D-$\sigma$ (size-velocity dispersion) correlation
for giant H~II regions 
has been firmly established (see \cite{M1977}, \cite{TM1981}, \cite{AR1988}),
but it is still not clear what this relationship means. 
L and D should correlate with mass; therefore one expects the velocity 
dispersion to also increase. However, the energy
input to the ISM in the form of ionizing photons, stellar winds, and 
supernovae should also correlate with L and D, increasing the amount
of ionized structure (expanding shells, filaments, etc.) and 
turbulent motions, both of which may broaden the integrated emission 
profiles.

There is evidence that apertures which 
include multiple ionized structures
at various velocities can create a smooth Gaussian
profile with a broadened core (\cite{CK1994}, \cite{Yetal1996}). In
a study of NGC 604, the brightest H II region in M33, Yang et
al. (1996) showed that although some of the broadening in the
integrated profile is due to averaging over many emitting regions, the
mean corrected FWHM still shows a substantial velocity dispersion
above that expected from purely thermal broadening. They concluded
that the extra 30 $\kms$ could be explained as virial motions. This is
in contrast to the case of the 30 Doradus giant H II region as
examined by Chu \& Kennicutt (1994), where even a tenfold increase in
the estimated mass of the region yielded a gravitational velocity
dispersion that was negligible in comparison with the observed
dispersion. In this case the conclusion was that the dominant contribution
to the global velocity dispersion in 30 Dor is from shell motions, but with
turbulence also playing a role.

It is not clear what is causing the width of the narrow emission cores 
in the case of
NGC~7673. If it is due to gravitational motions, we should expect to
see wider lines in connection with the more massive H II regions as
shown in the H$\alpha$ image. Broadening due to averaging over many
bubbles and filaments should also produce a spatial variation in the
line widths, as some fibers cover a greater number of H~II regions than 
others. Due
to the low spatial resolution of our observations and the uncertainty
in fiber placement, we cannot rule out either of these
possibilities. The observed scatter of approximately 10 km s$^{-1}$in the 
total FWHM about the
mean value implies an extra velocity component of width $\sim$ 35 km s$^{-1}$ 
(FWHM, or $\sigma \sim$ 15 km s$^{-1}$) is possible.  

We also see a broad emission component underlying the (relatively)
narrow cores, which have been previously observed in H$\alpha$ 
emission lines, but are not well-understood. 
Broadened cores produced by averaging over many emitting areas have been
found accompanied by  
low-intensity wings (\cite{CK1994}), presumably from expanding shells
and bubbles. The recent study of NGC 604 (\cite{Yetal1996}) also
showed broad wings in the integrated profiles, again, from the
inclusion of many expanding shells. It would be interesting to see if
these integrated profiles can be deconvolved into narrow and broad
components, and how the width and/or shape of the lines change as one 
averages over a larger area.

There are three leading explanations for the broad wings: they may be the 
result of integrating over many ionized structures at different velocities, 
or a broad emission component may be present from hot, turbulent gas confined 
to large cavities carved out by massive stars, or they could be due to some 
type of break-out phenomenon, such as a 
champagne flow or a starburst-powered galactic wind.

Massive outflow phenomenon is highly unlikely because of the 
absence of double-peaked emission profiles and/or a substantial velocity 
offset between the narrow and broad components, both of which are 
characteristic 
of superwinds (\cite{HAM1990}, \cite{LH96}). This leaves us with the first 
two explanations as listed above. Given the results of the studies by 
Chu \& Kenicutt (1994) and Yang et al. (1996), we favor the first 
explanation, where the line profiles are the result of integrating over 
many shells, bubbles, and filaments, which can also explain the highly
supersonic line widths of the narrow component.

\subsection{Starburst Trigger}

NGC~7673 is a close projected companion to NGC~7677, which has a
similar radial velocity; therefore these two galaxies are probably a
physical pair (\cite{CH1976}, \cite{Netal1997}).  The surrounding 
area is free of
small H~I sources. A few minor appendages to the main disk of NGC~7673
are seen in the low-resolution H I map of Nordgren et at. (1997), but 
there is an absence of
classic interaction signatures, such as H~I tails. Given the disturbed
optical appearance of NGC~7673, its overall H~I morphology is surprisingly
symmetric. No H~I peculiarities near the location of the outer optical shell
were detected, although this may be due to the low resolution of the H~I map. 
As mentioned previously, there is a warp in the disk of NGC~7677 and an H~I
extension pointing towards NGC~7673, but the outer H~I envelopes are
widely separated. The H~I structure in these galaxies displays no
indications of a significant on-going interaction between the two; any serious
interaction would need to have taken place long enough ago for the outer
H I disks to recover to their observed relatively normal states.

The discrepancy between the morphology of NGC~7673 and NGC~7677 is
apparent on the Digital Sky Survey\footnote{The Digitized Sky Survey
was produced at the Space Telescope Science Institute under
U.S. Government grant NAG W-2166. The images of these surveys are
based on photographic data obtained using the Oschin Schmidt Telescope
on Palomar Mountain and the UK Schmidt Telescope. The plates were
processed into the present compressed digital form with the permission
of these institutions.}  frames; NGC~7677 looks relatively regular, 
with a
bright nuclear region and grand design spiral arms, in dramatic contrast to
NGC~7673, whose high levels of star formation create giant H II
regions and a remarkably clumpy appearance. If the
starburst in NGC~7673 is due to an interaction with NGC~7677, then the
interaction scenario must be able to explain why NGC~7673 is so
disturbed, and NGC~7677 relatively unscathed. We conclude that an on-going 
interaction with NGC~7677 is not the starburst trigger. Even though no strong 
collision is currently in progress between NGC~7673 and NGC~7677, the 
disturbed outer optical disk of NGC~7673 does seem to be the signature of a
past interaction.

The sharp outermost
arc, along with the wispy extensions and faint ripples that surround
the bright inner optical disk, are features usually associated with
merger candidate E, S0, and a few Sa galaxies (\cite{SS1988}). These 
structures are
rare in later-type galaxies, but have also been found around another
spiral starburst galaxy, NGC~3310 (\cite{MvD1996}). There are striking
similarities between NGC~7673 and NGC~3310. They are both starburst
galaxies with inner spiral structure surrounded by faint arcs and
ripples, and their H$\alpha$ emission lines are strong and asymmetric
(\cite{GS1991}). The general consensus is that NGC~3310 has gone
through a minor merger with a dwarf companion, which accounts for
the starburst, peculiar arc, and ripple features in its outer parts
(\cite{GS1991}, \cite{MvD1996}, \cite{Setal1996},
\cite{MvDB1995}, \cite{Ketal1993}).  The minor merger model therefore
is also attractive for NGC~7673.

Theoretical work concerning ripples and 'shells' around galaxies has
shown that they can be formed in a variety of circumstances around
both elliptical and disk galaxies (\cite{Q1984}, \cite{WS1988}, \cite{HQ1989}, 
\cite{HS1992}, \cite{H93}). Observationally, it is often difficult to 
distinguish
between the models, and thus the origin of ripples in any particular object
is often a subject of 
debate. Thus while mergers are the preferred model for the production of
ripples and related features, their origin in other types of interactions
cannot be excluded. 

We therefore have two possible models: the NGC~7673 
starburst could have been triggered by a past interaction with NGC~7677 that 
occurred long enough ago that any tidal streamers are gone, or a small galaxy
could have merged with NGC~7673 in the past and the nearby presence of 
NGC~7677 is simply fortuitous. In either case the collisional event must have
occurred long enough ago to allow the outer disk to mostly recover, and have
been mild enough to avoid serious disruption of the main disk. Any interloper
must therefore have been much  smaller than the accreting galaxy, 
probably $<$ 10 $\%$ of the mass of the disk (\cite{HM1995}),
making NGC~7673 a minor merger candidate.

\section{Summary and conclusions}

We have obtained moderate resolution optical images and echelle spectroscopy
of the luminous blue galaxy NGC~7673. Our B-R color map shows red dust lanes 
and blue star
clusters in an irregular spiral pattern, and the H$\alpha$ image confirms 
that these bright clusters
are also giant H II regions. The deep B and R -band images confirm the
presence of the previously reported ripple 1'.55 west of the galaxy
center (\cite{Detal1984}), as well as revealing several other faint
extensions and ripples surrounding the bright optical disk. These features
are composed of stars, probably in the outer disk of NGC~7673.

Our fiber array measurements of the H$\alpha$ kinematics demonstrate that 
the H~II regions are embedded in
a smoothly rotating disk that we are viewing near a rotational
pole. Despite the presence of a large-scale starburst, the disk
velocity field appears to be remarkably regular. In areas of low flux,
the emission profiles are often highly asymmetric, and are not well
described by a single Gaussian profile. We may be seeing line
splitting from isolated ionized super-shells, or the lines may be a 
combination of
two or more H~II regions with different radial velocities. Double
Gaussian fits were performed on emission lines from the central
regions of the galaxy to investigate the nature of the broad wings of
emission. We find that a two component model, one narrow and one
broad, fit the observed spectra rather well. The narrow component of complex
emission line profiles have 
the same width as the single component lines, FWHM $\sim$ 55 $\kms$,
and the broad component has a FWHM $\sim$ 150 $\kms$. The broad lines
often have a slight velocity offset ($<$ 10$\kms$), blue-ward of the
narrow lines.

We can exclude an active interaction with NGC 7677 as the starburst trigger,
but we cannot rule out a past interaction. If NGC 7677 is the culprit, then
the starburst phase must be sustained well after the main encounter. The other
possibility is capture of a dwarf companion, as in the case of NGC 3310, the 
leading candidate for a major starburst induced by a minor merger. The 
morphological similarities between NGC~7673 and NGC~3310 and NGC~7673's 
 symmetric outer H I disk (\cite{Netal1997}) lend support to the minor
merger hypothesis. High resolution H~I observations may resolve this issue.

\acknowledgements It is a pleasure to thank the many people who make
the WIYN 3.5-m an outstanding research telescope, and especially Sam
Barden and Dave Sawyer for bringing DensePak online. We would like to thank 
Chris Dolan and Chris Anderson for their help with DensePak data reduction. 
We would also like to thank an anonymous referee for comments which improved 
the paper and for pointing out the possibly rather isolated location of 
NGC~7673. This project is a component of WFPC2
Investigation Definition Team studies of luminous blue galaxies, and
is supported in part by NASA contract NAS 7-1260 to the Jet Propulsion
Laboratory.

\newpage

\figcaption[figure1.ps]{The DensePak fiber array: each fiber has a 3'' 
diameter; fiber centers are separated by 4'', fiber numbers are also shown.}
   
\figcaption[figure2.ps]{WIYN CCD images, north is up and east is to the left, 
(a)-(c) are R band: (a.) diffuse extensions (b.) outer arcs or 'ripples' (c.) 
inner structure (d.) B/R color map: red (black) dust lanes and blue (white) 
star clusters; Note the background galaxies are in the upper right and 
confused in the condensation to the southeast.}

\figcaption[figure3.ps]{Continuum subtracted H$\alpha$ image, the box 
corresponds to placement of the DensePak array.}

\figcaption[figure4.ps]{DensePak map, flux in grayscale. Fiber 39 is marked 
for orientation. North is up and east is to the left.}

\figcaption[figure5.ps]{Examples of spectra with varying signal-to-noise, 
fiber number is at the top, solid line is the spectrum, dotted line is the 
fit. Note the broad wings in the fiber 57 spectrum.}

\figcaption[figure6.ps]{Position-FWHM diagram illustrating the lack of spatial 
variation in narrow line width. Plotted FWHMs are corrected for the 
32 km s$^{-1}$ instrumental broadening determined from night sky lines. 
Position is measured from the center of the 
southern-most row of fibers. The figure was made by plotting the FWHM values 
in each fiber along rows, all fibers in a row are at the same position value. 
The mean FWHMs of the two components are 56 and 149 $\kms$. Due to large 
asymmetries and/or low S/N, we have excluded fibers 16, 33, 41, 42, 48, 55, 
62, 72, 75, 79, 80, and 81.}

\figcaption[figure7.ps]{Heliocentric velocities of the narrow, FWHM $\sim$ 55 
$\kms$, lines, in km s$^{-1}$. Fibers with no emission are colored black, and
grayscale shading is included to aid visualizing the velocity field.}


\newpage

\begin{thebibliography}{sbsts}

\bibitem[Arp 1965]{A66} Arp, H. 1966, ApJ Supp. Ser. 14, 1

\bibitem[Arsenault \& Roy 1986]{AR1986} Arsenault, R., \& Roy, J.-R. 1986, AJ, 92, 567

\bibitem[Arsenault \& Roy 1988]{AR1988} Arsenault, R., \& Roy, J.-R. 1988, A\&A, 201, 199

\bibitem[Barden et al. 1998]{BSH98} Barden, S., Sawyer, D., \& Honeycutt, R. 1998, Proc. SPIE, 3355, p.892-899

\bibitem[Benvenuti et al. 1982]{BCH1982} Benvenuti, P., Casini, C., Heidmann, J. 1982, MNRAS, 198, 825

\bibitem[B\"{o}rngen \& Kalloglian 1975]{BK1975} B\"{o}rngen, F., Kalloglian, A. 1975, ApJ, 11, 24

\bibitem[Casini \& Heidmann 1976]{CH1976} Casini, C., \& Heidmann, J. 1976, A\&A, 47, 371

\bibitem[Chu \& Kennicutt 1994]{CK1994} Chu, Y., \& Kennicutt, R. 1994, ApJ, 425, 720

\bibitem[Coupinot et al. 1982]{Cetal1982} Coupinot, G., Hecquet, J.,  \& Heidmann, J. 1982, MNRAS, 199, 451

\bibitem[Cowie, Hu, \& Songalia 1995]{CHS1995} Cowie, L., Hu, E., Songalia, A. 1995, AJ, 110, 157

\bibitem[Dettmar et al. 1984]{Detal1984} Dettmar, R., Heidmann, J., Klein, U., \& Wielebinski, R. 1984, A\&A, 130, 424

\bibitem[de Vaucouleurs, de Vaucouleurs, \& Corwin 1976]{DDC1976} DeVaucouleurs, G., DeVaucouleurs, A., Corwin, H. 1976, 2nd Ref. Cat. of Bright Galaxies, U of Texas Press, Austin, TX

\bibitem[Duflot-Augarde \& Alloin 1982]{DA1982} Duflot-Augarde \& Alloin, D. 1982 A\&A, 112, 257

\bibitem[Fich, Treffers, \& Dahl 1990]{FTD90} Fich, M., Treffers, R., Dahl, G. 1990, AJ, 99, 622

\bibitem[Ellis 1997]{E1997} Ellis, R. 1997, ARAA, 35, 389

\bibitem[Gallagher 1990]{G1990} Gallagher, J. 1990, in Evolution of the Universe of Galaxies, ASP 
Conf Series 10, ed. R. G . Kron, p157 

\bibitem[Gallagher \& Hunter 1983]{GH1983} Gallagher, J., Hunter, D. 1983, ApJ, 274, 141

\bibitem[Gallagher, Hunter, \& Bushouse 1989]{GHB89} Gallagher, J., Hunter, D., Bushouse, H. 1989, AJ, 97, 700

\bibitem[Gallagher \& Gibson 1994]{GG94} Gallagher, J. S. \& Gibson, S. J. 1994, in Panchromatic View of Galaxies - Their Evolutionary Puzzle, eds. G. Herster, Ch. Theis, J. Gallagher (Gif-sur-Yvette:Editions Frontiers), p.207 

\bibitem[Gallagher et al. 1999]{Getal99} Gallagher, J., Homeier, N., Griffiths,R., \& WFPC2 IDT 1999, in preparation

\bibitem[Gallego et al. 1996]{Getal1996} Gallego, J., Zamorano, J., Rego, M., Alonso, O., Vitores, A. G. 1996, A\&AS,  120, 323

\bibitem[Grothues \& Schmidt-Kaler 1991]{GS1991} Grothues, H., Schmidt-Kaler, Th. 1991, A\&A, 242, 357

\bibitem[Guzman et al. 1998]{Getal98} Guzman, R., Jangren, A., Koo, D., Bershady, M., Simard, L. 1998, ApJ, 495, L13

\bibitem[Heckman, Armus, \& Miley 1990]{HAM1990} Heckman, T., Armus, L., Miley, G. 1990, ApJS, 74, 833

\bibitem[Hernquist \& Mihos 1995]{HM1995} Hernquist, L., \& Mihos, J. 1995, ApJ, 448, 41

\bibitem[Hernquist \& Quinn 1989]{HQ1989} Hernquist, L., Quinn, P., 1989, ApJ, 342, 1

\bibitem[Hernquist \& Spergel 1992]{HS1992} Hernquist, L., \& Spergel, D. 1992, ApJ Letters, 399, 117

\bibitem[Howard et al. 1993]{H93} Howard, S., Keel, W.C., Byrd, G., \& Burkey, J. 1993, ApJ, 417, 502

\bibitem[Huchra 1977]{H77} Huchra, J. 1977, ApJS, 35, 171

\bibitem[Hunter et al. 1994]{dah94} Hunter, D. A., van Woerden, H., \& Gallagher, J. S. 1994, ApJS, 91, 79

\bibitem[Lehnert \& Heckman 1996]{LH96} Lehnert, M., Heckman, T. 1996, ApJ 472, 546

\bibitem[Kikumoto et al. 1993]{Ketal1993}Kikumoto, T., Taniguchi, Y., Suzuki, M., \& Tomisaka, K. 1993, AJ, 106, 2, 466

\bibitem[Markarian \& Lipovetski 1971]{ML1971} Markarian, B., \& Lipovetski, V. 1971, Astrofisika, 7, 511

\bibitem[Melnick 1977]{M1977} Melnick, J. 1977, ApJ, 213, 15

\bibitem[Mendez \& Esteban 1997]{ME1997} Mendez, D., Esteban, C. 1997, ApJ, 488, 652

\bibitem[Mihos \& Hernquist 1994]{MH1994} Mihos, J., Hernquist, L. 1994, ApJ, 425, L13

\bibitem[M\"{u}nch 1958]{M58} M\"{u}nch, G. 1958, Rev. Mod. Phys., 30, 1035

\bibitem[Mulder \& van Driel 1996]{MvD1996} Mulder, P., van Driel, W. 1996, A\&A, 309, 403

\bibitem[Mulder et al. 1995]{MvDB1995} Mulder, P., van Driel, W., Braine, J. 1995, A\&A, 300, 687

\bibitem[Nordgren et al. 1997]{Netal1997} Nordgren, T., Chengalur, J., Salpeter, E., \& Terzian, Y. 1997, AJ, 114, 77

\bibitem[Quinn 1984]{Q1984} Quinn, P., 1984, ApJ, 279, 596

\bibitem[Roy, Arsenault, \& Joncas 1986]{RAJ1986} Roy, J., Arsenault, R., Joncas, G. 1986, ApJ, 300, 624

\bibitem[Rozas et al. 1998]{Retal1998} Rozas, M., Sabalisck, N., Beckman, J., Knapen, J. 1998, A\&A, 338, 15

\bibitem[Sandage 1963]{S63} Sandage, A. 1963, ApJ, 138, 863

\bibitem[Sargent 1970]{S1970} Sargent, W., 1970, ApJ, 160, 405

\bibitem[Schweizer \& Seitzer 1988]{SS1988} Schweizer, F., Seitzer, P. 1988, AA, 328, 88

\bibitem[Skillman \& Balick 1984]{SB1984} Skillman, E., \& Balick, B. 1984 ApJ, 280, 580

\bibitem[Smith \& Weedman 1970]{SW1970} Smith, M., \& Weedman, D. 1970 ApJ, 161,33

\bibitem[Smith et al. 1996]{Setal1996} Smith, D., Neff, S., Bothun, G., Fanelli, M., Offenberg, J., Waller, W., Bohlin, R., O'Connell, R., Roberts, M., Smith, A., \& Stecher, T. 1996, ApJ, 473, L21

\bibitem[Tamura \& Heidmann 1986]{TH1986} Tamura, S., Heidmann, J. 1986, Publ. Astron. Soc. Japan, 38, 619

\bibitem[Taniguchi \& Tamura 1987]{TT1987} Taniguchi, Y., \& Tamura, S., 1987, A\&A, 181, 265

\bibitem[Terlevich \& Melnick 1981]{TM1981} Terlevich, R., \& Melnick, J. 1981, MNRAS, 195, 839

\bibitem[Walker et al. 1996]{Wetal1996} Walker, I., Mihos, J., \& Hernquist, L. 1996, ApJ, 460, 121

\bibitem[Wallin \& Struck-Marcell 1988]{WS1988} Wallin, J., Struck-Marcell, C. 1988, AJ, 96, 1850

\bibitem[Yang et al. 1996]{Yetal1996} Yang, H., Chu, Y.-H., Skillman, E., Terlevich, R. 1996, AJ, 112, 146 

\bibitem[Zwicky 1957]{Z57} Zwicky, F. 1957, Morphological Astronomy (Berlin: Springer Verlag).

\end{thebibliography}
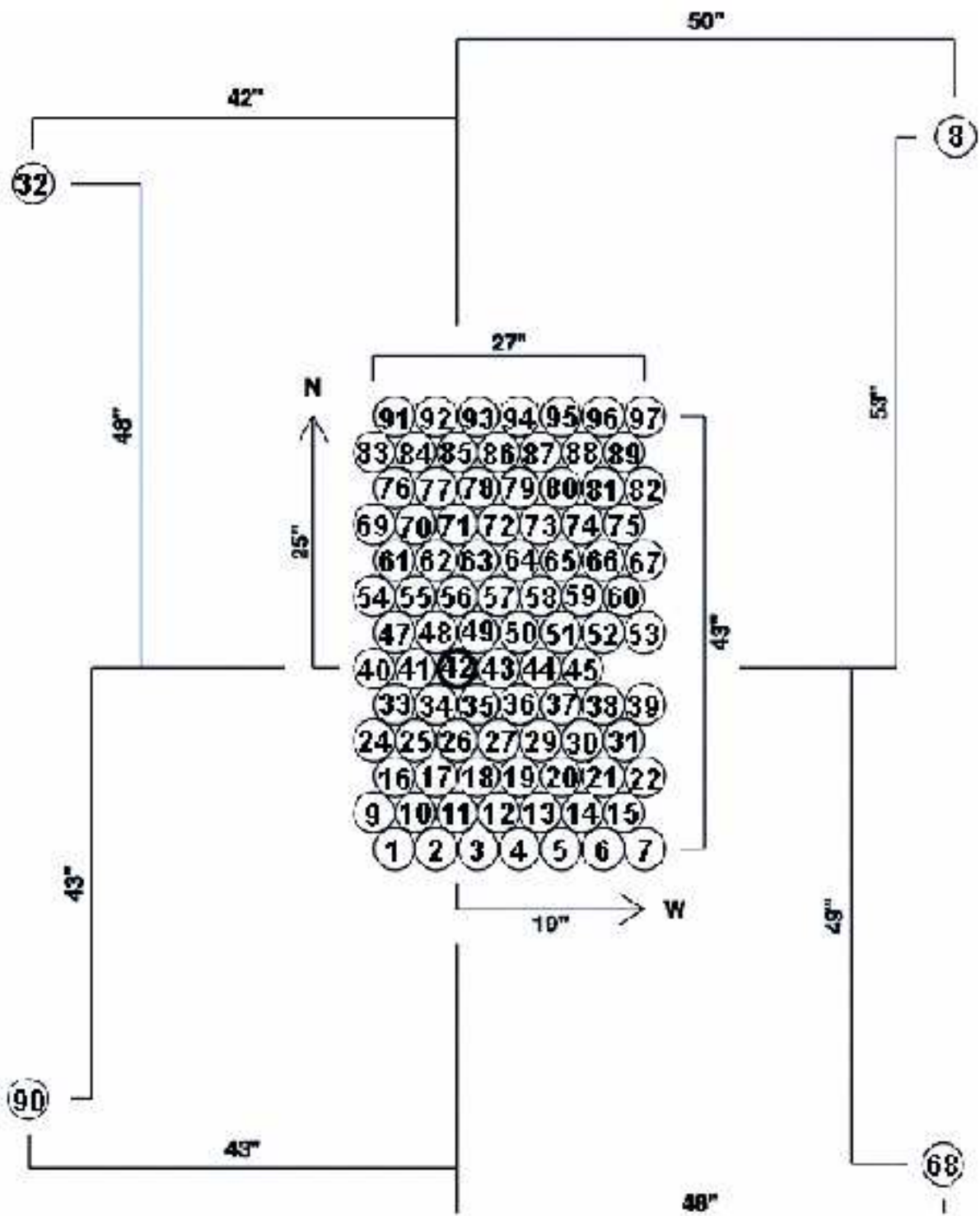
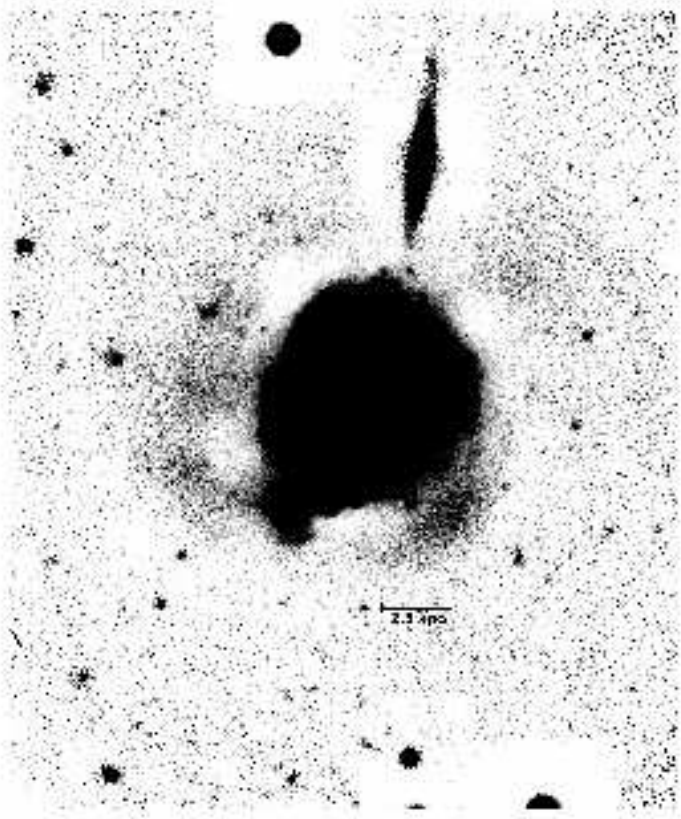
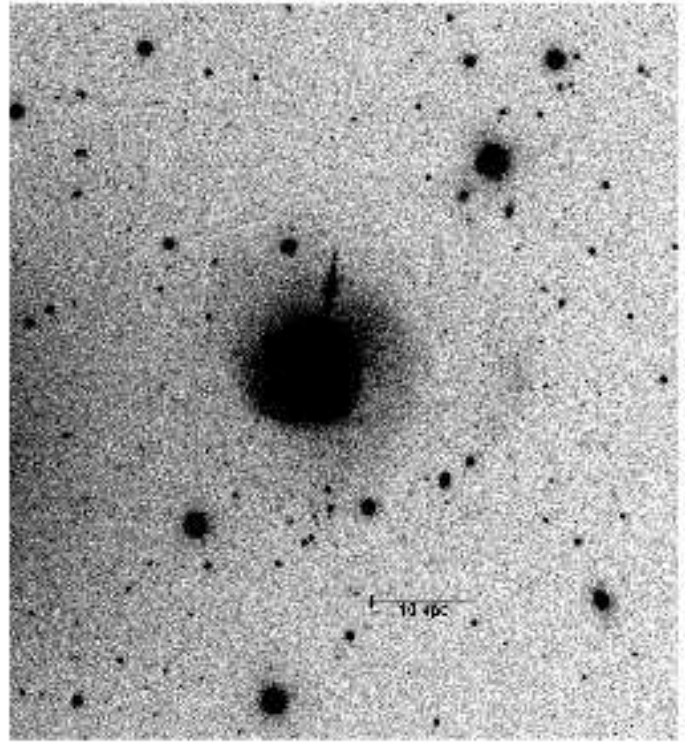
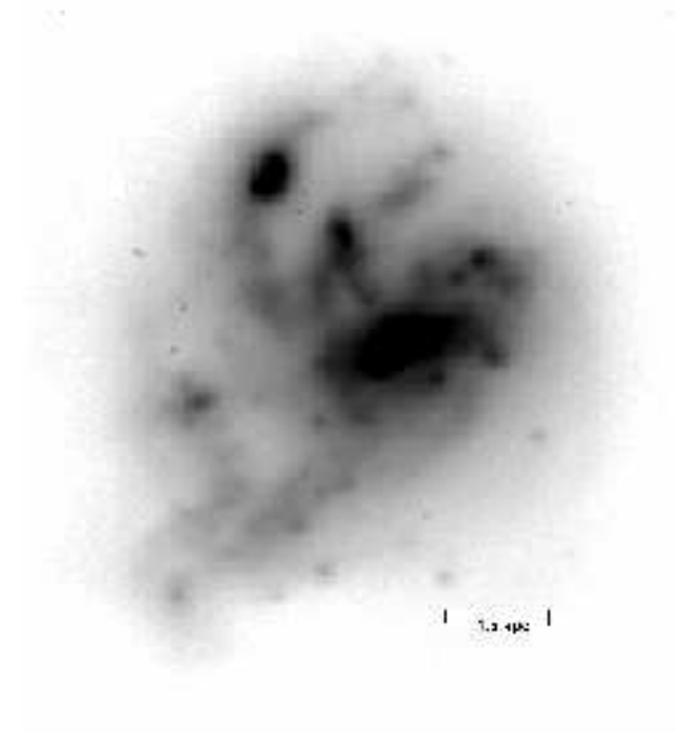
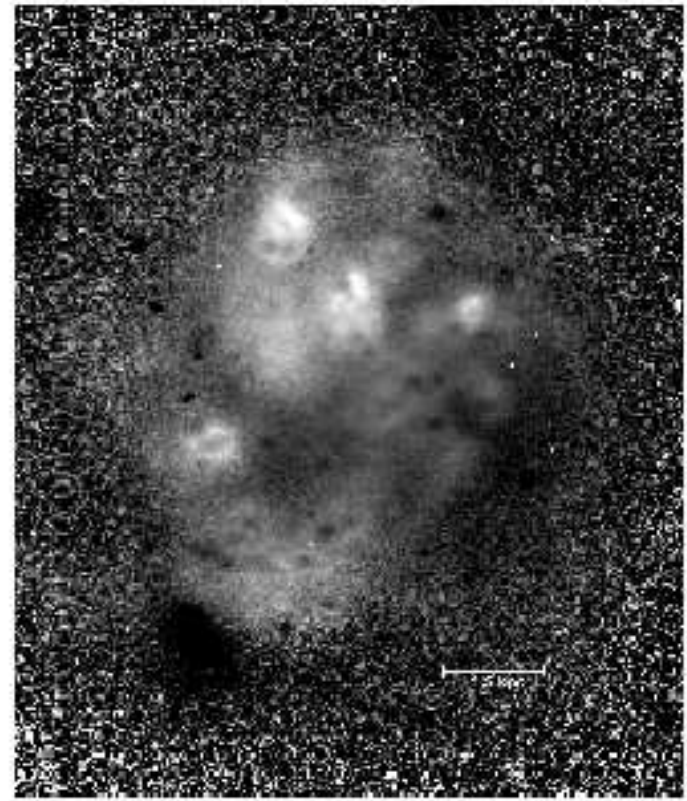
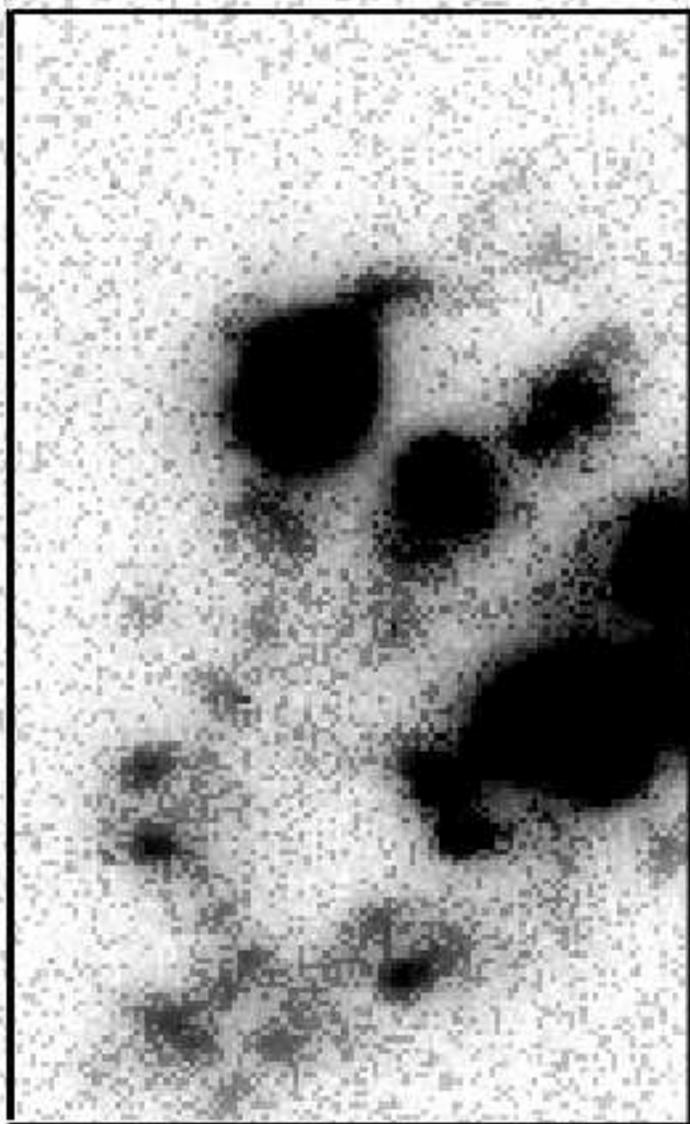
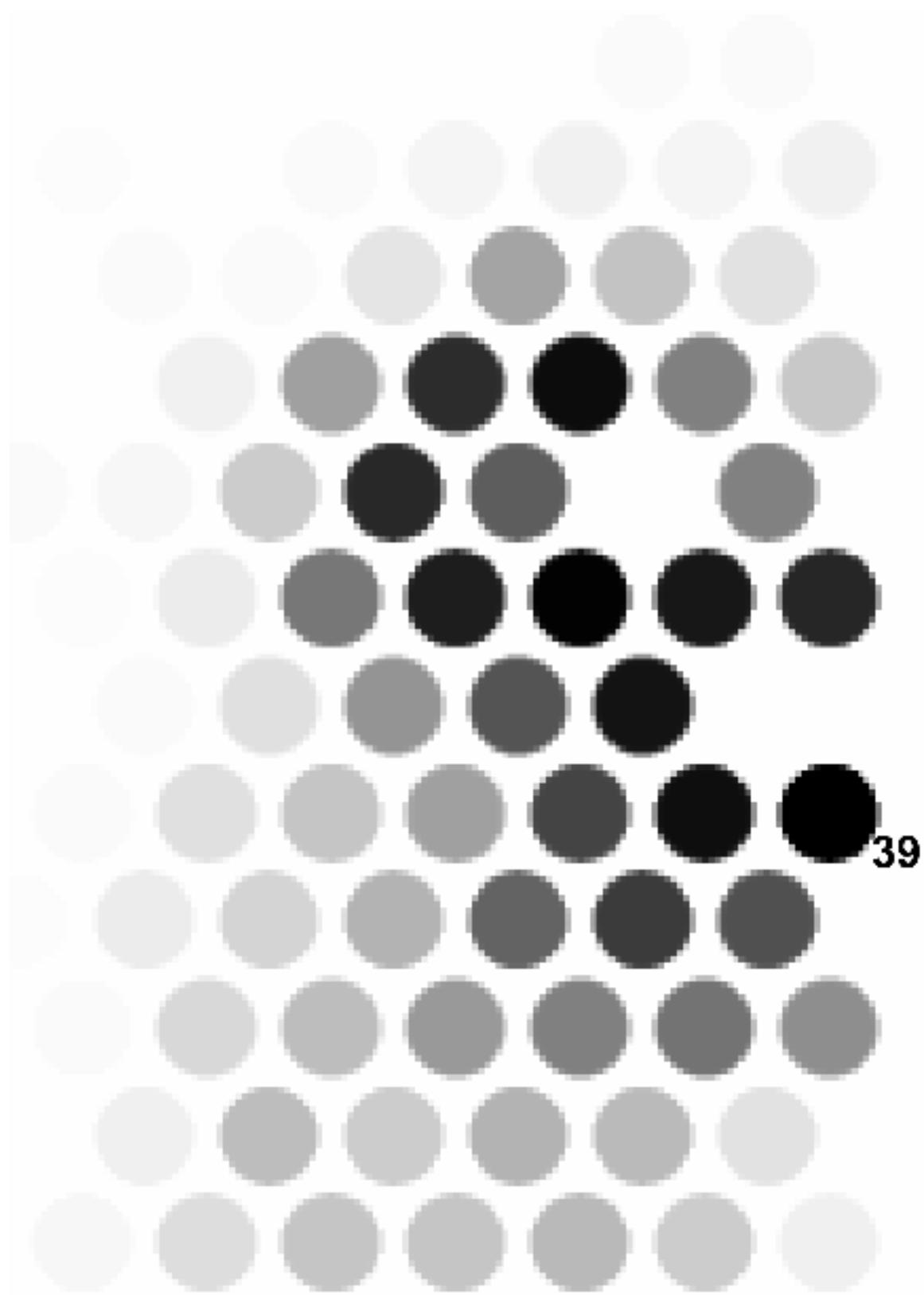
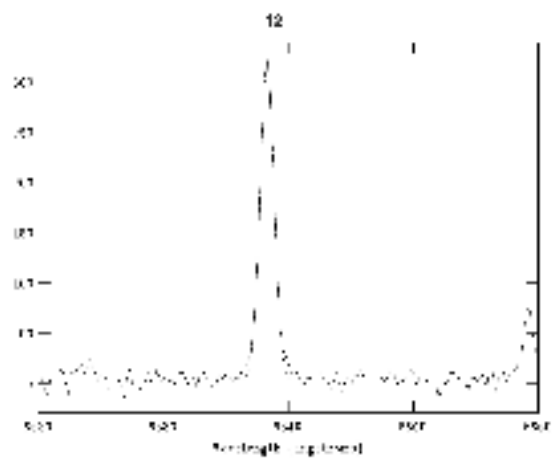
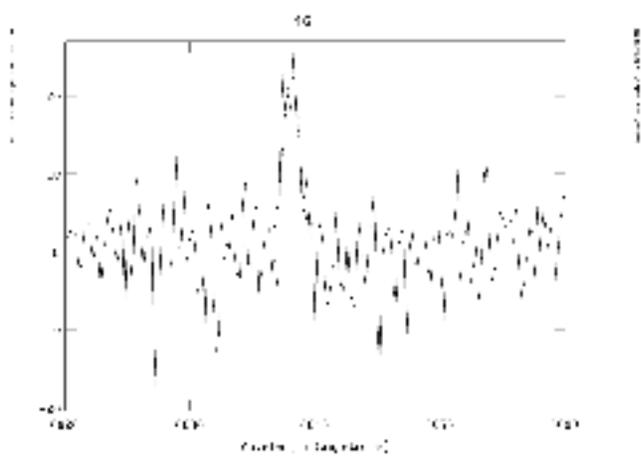
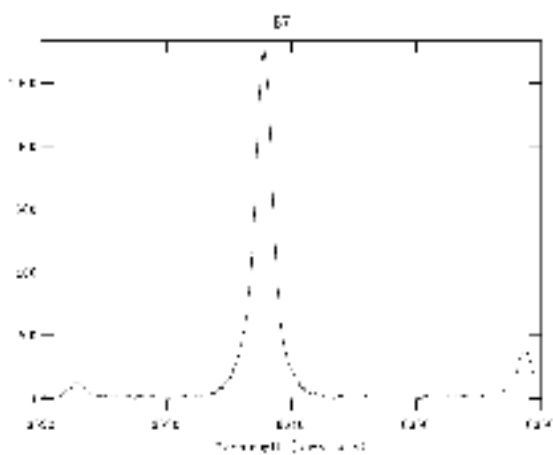
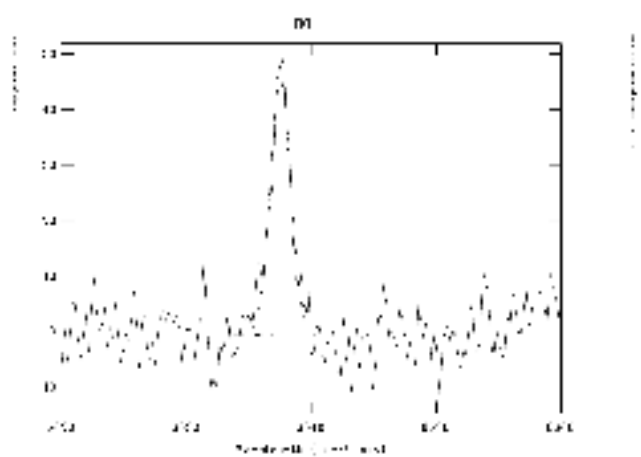
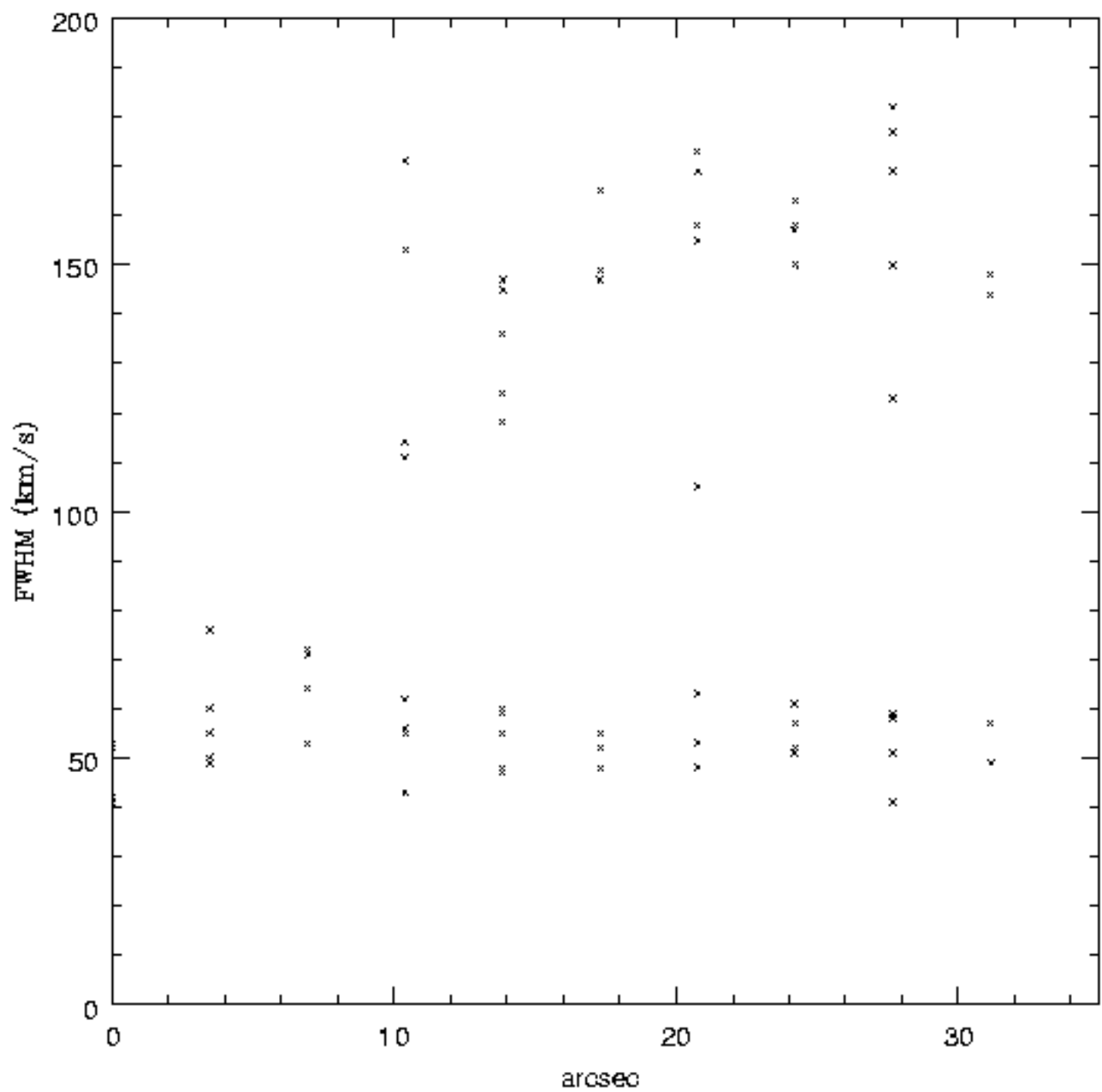
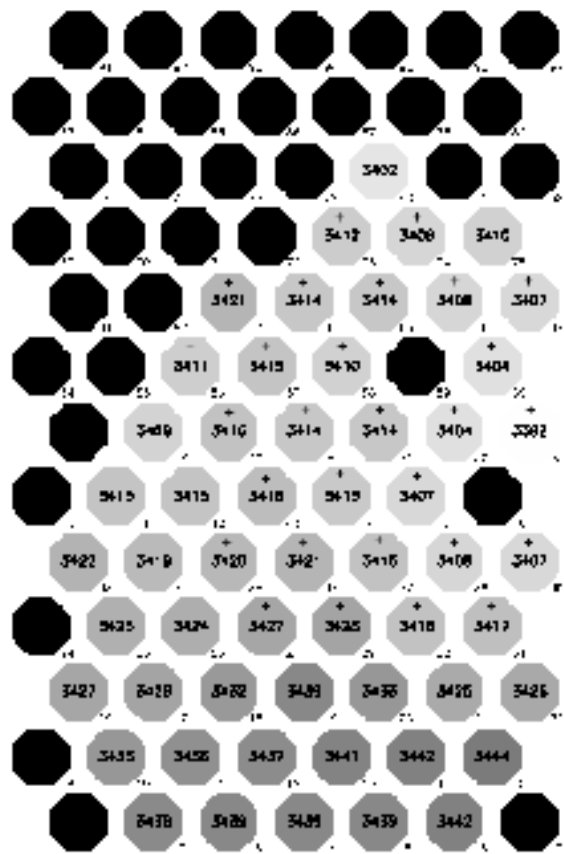
\end{document}